# A Compact Neutron Scatter Camera for Field Deployment

John E. M. Goldsmith,[1] Mark D. Gerling, and James S. Brennan

Sandia National Laboratories; MS 9406, PO Box 969; Livermore, CA 94551; USA

We describe a very compact (0.9 m high, 0.4 m diameter, 40 kg) battery operable neutron scatter camera designed for field deployment. Unlike most other systems, the configuration of the sixteen liquid-scintillator detection cells are arranged to provide omnidirectional ($4\pi$) imaging with sensitivity comparable to a conventional two-plane system. Although designed primarily to operate as a neutron scatter camera for localizing energetic neutron sources, it also functions as a Compton camera for localizing gamma sources. In addition to describing the radionuclide source localization capabilities of this system, we demonstrate how it provides neutron spectra that can distinguish plutonium metal from plutonium oxide sources, in addition to the easier task of distinguishing AmBe from fission sources.

## I. Introduction

We have developed and deployed a very compact and mobile neutron detection system for finding (localizing) and characterizing fission-energy (~1-10 MeV) neutron sources, with the primarily envisioned application being use by emergency responders. This system builds on the decade-plus work we have performed on developing neutron scatter cameras[1-3] that led to our current large segmented imaging system optimized for large-area search applications. The system described in detail here has been engineered to fit a much smaller form factor, and to be operated by either a battery or AC power.[4] Because the direction to a neutron source may not be known *a priori*, we chose a design that enabled omnidirectional ($4\pi$) imaging, with only a ~twofold decrease in sensitivity compared to the much larger neutron scatter cameras (at the cost of reduced spatial and spectral resolution). The system design was optimized for neutron imaging and spectroscopy, but it also functions as a Compton camera for gamma imaging. This paper describes the system in detail, and presents laboratory measurements made with it.

---

[1] Author to whom correspondence should be addressed. Electronic mail: jgold@sandia.gov

## II. Imaging methodology

Several approaches have been demonstrated for localizing neutron sources. For some applications, the straightforward approach of walking around with a neutron counter until the source is found is suitable. If the application calls for stand-off localization, however, an imaging technique is needed. In addition to the neutron-scatter-camera approach described here and elsewhere,[1-11] other approaches include collimated-detector scanning systems,[12] coded aperture systems,[13] and one-d and two-d time-encoded imaging systems.[14,15] All approaches have their advantages, and the choice depends on the details of the application. The neutron scatter camera offers modest sensitivity and spatial resolution (appropriate for localization, not true imaging) and excellent gamma-background rejection over a very large field-of-view ($4\pi$ for the design described here) with a moderate channel count, moderate size, and no moving parts.

Scatter-camera imaging of fast neutrons utilizes the kinematics of elastic scattering to reconstruct the incident energy and direction of neutrons that interact twice in a detector system as illustrated in Figure 1. The incident neutron scatters off a proton in the first detector, and the deposited energy $E_{D1,N}$ is determined from the scintillation signal in that detector. If the scattered neutron is subsequently observed in a second detector, the energy $E_{S,N}$ of the scattered neutron is determined by the time-of-flight $\tau$ between the two detectors separated by distance d (Eqn. 1). (The velocity of a 1 MeV neutron is ~5% c, hence the use of this nonrelativistic formula.) The energy of the incident neutron is the sum of the energy deposited in the first cell and the energy of the scattered neutron (Eqn. 2). The kinematic constraint imposed by those two energies determines the initial scatter angle $\theta$ shown in the figure to within the surface of the cone with the angle $\theta_N$ derived from Eqn. 3.[16] For neutrons scattering off protons, this angle cannot be greater than 90°. For back-projection image reconstruction, the intersection of the cones as measured by multiple neutron events then determines the direction to the source. It is worth noting that the signal from the second cell is only used to derive the time-of-flight $\tau$ of the scattered neutron; it cannot be used to accurately determine the energy of the scattered neutron as the deposited energy can be anywhere between a very small value and the full energy of the neutron.

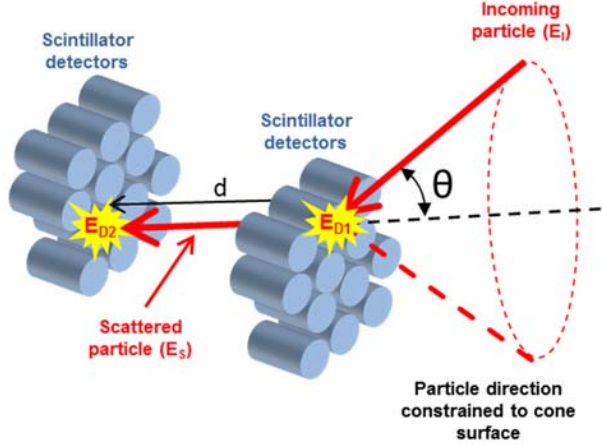

Fig. 1. Scattering geometry and quantities relevant to image reconstruction

$$E_{S,N} = \frac{1}{2} m \left(\frac{d}{\tau}\right)^2 \tag{1}$$

$$E_{I,N} = E_{D1,N} + E_{S,N} \tag{2}$$

$$\theta_N = \sin^{-1} \sqrt{E_{D1,N}/E_{I,N}} \tag{3}$$

Compton-camera imaging of gammas is performed in a similar manner, but we cannot use the time-of-flight of the scattered gamma to infer its energy. In a conventional Compton camera, the second detector is designed to capture and measure the full energy of the scattered gamma. This is not the case for the low-Z, modest-size liquid-scintillator cells employed in this system. In general, it is necessary to derive an estimate for the fraction of the energy of the scattered gamma ($E_{S,G}$) that is measured in the second cell ($E_{D2,G}$). To derive this value, we utilize an MCNP-PoliMi simulation of similar 3" x 3" liquid scintillator cells and various input gamma energy spectra. The results are analyzed event-by-event in post processing to reconstruct energy deposited, before being smeared as a function of energy and compared to measured data in order to extract the energy calibration scale. We have made use of these energy deposited spectra from simulation in order to extract the average energy a gamma event will deposit based on its initial energy. Due to Compton edge, down scattering, and other edge effects, this will always be less than the full energy of the event. By integrating the simulation results above a threshold and comparing the energy deposited to the total energy available, we were able to extract the fraction energy deposited for our cells at each energy. In practice, we have observed that with the operational 0.1 MeV threshold of the system, it is 49% for 0.662 and 52% for 1.275 MeV gamma events. This has allowed us to extrapolate a corrective factor of 2 for gamma imaging in this energy range, which very closely agrees with measured results and calibrations. The energy of the incident gamma is the sum of the energy deposited in the first cell and the energy of the scattered gamma (Eqn. 5). The scatter angle $\theta_G$ is then given by Eqn. 6 (where $m_e$ is the mass of the electron).[17]

$$E_{S,G} \approx 2 * E_{D2,G} \tag{4}$$

$$E_{I,G} = E_{D1,G} + E_{S,G} \tag{5}$$

$$\theta_G = \cos^{-1}\left(1 - \frac{E_{D1,G} * m_e c^2}{E_{I,G} * E_{S,G}}\right) \tag{6}$$

### III. System configuration

Many choices go into the design of a neutron scatter camera. Single-volume designs have been proposed and are under investigation, but we restrict ourselves here to designs that use multiple detection cells. Design choices related to gamma discrimination, sensitivity, spatial resolution, and spectral resolution must be traded off with practical system constraints (size, weight, power requirements, and cost). The need for gamma discrimination in order to detect weak neutron emissions in possibly strong gamma backgrounds suggests the use of a scintillator with pulse-shape-discrimination (PSD) capabilities. Until very recently, this meant liquid scintillator detectors, with the non-hazardous, high-flash-point material EJ-309 being chosen to avoid difficulties in transporting the system. The recent commercial availability of PSD-capable plastic scintillator and Stilbene crystals lead to additional material choices, but do not significantly affect the other design choices related to performance. The desire for high sensitivity suggests a large volume of scintillator material, with close spacing of detection cells to optimize the probability of double-scatter events. The larger the cell size, the higher the probability of a neutron interaction in the cell (Fig. 2). However, our image analysis assumes that only a single neutron-scatter event occurs in each cell; the larger the cell, the worse this assumption. The desire for high spatial and spectral resolution also suggests small cells (to better define the interaction location) and large separation (for accurate time-of-flight measurements). Our requirement for a compact, lightweight, low-power, relatively inexpensive system suggests minimal detector volume and cell count in a close-spaced configuration. Within the design space compatible with those requirements, we chose to optimize detection sensitivity at some loss in spatial and spectral resolution, and thus selected a close-spaced configuration with a modest number of fairly large cells. With a design goal of a system weight under 45 kg and after considering the available selection of fast (for PSD capability), sensitive photomultiplier tubes and appropriate commercially available fast multichannel digitizers, we chose to use sixteen 7.6 cm diameter liquid scintillator cells.

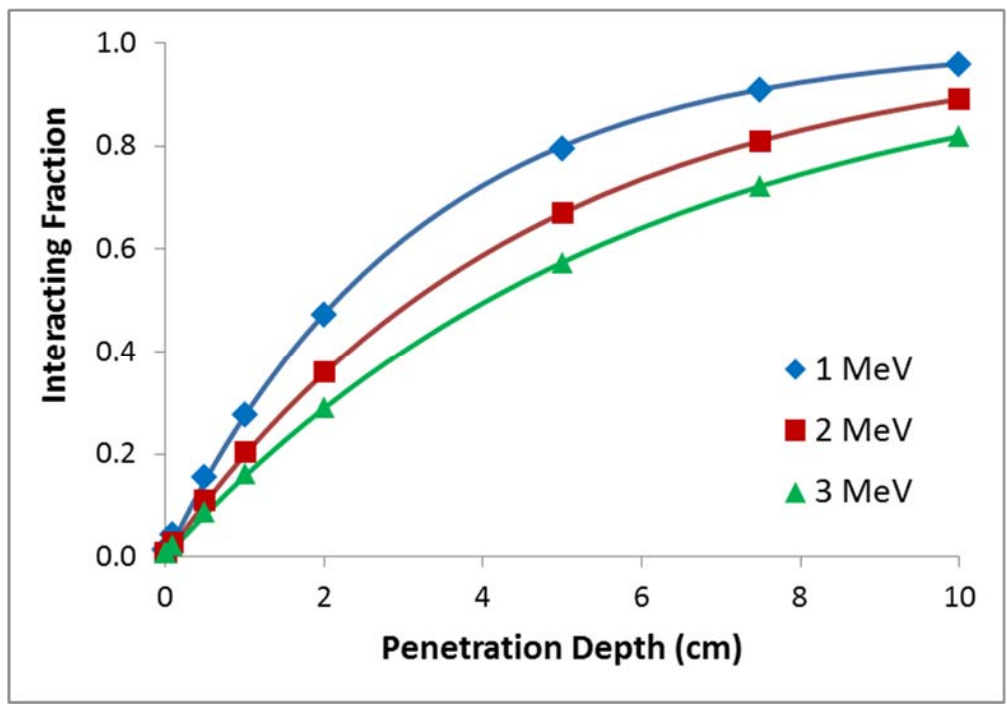

Fig. 2. Fraction of neutrons that interact with the liquid scintillator by the indicated penetration depth (data points: calculated using MCNP; solid curves: exponential fits to the data points).

Typical neutron scatter cameras and Compton cameras employ a two-plane design such as that shown in Fig. 1.[1-3,6] The two-plane approach is conceptually straightforward, but if the detection algorithm requires a scatter in each plane, the dominant system field-of-view is limited to two general directions. If the electronics are configured to allow it, either plane can serve as front or rear plane. We envision emergency-response measurement scenarios in which the direction of the source is not known and thus a system with omnidirectional sensitivity is preferred. We modeled the response of a large number of sixteen-detector configurations, ultimately choosing the height-offset cylindrical configuration shown in Fig. 3. This design provides very uniform sensitivity in the horizontal plane, with less than a factor-of-two variation in the vertical plane. The close spacing of the cells leads to sensitivity in all directions as good as a two-plane system with a plane separation equal to the radius of the cylindrical configuration. This configuration only has a ~twofold lower double-scatter efficiency than provided by our large neutron scatter camera with thirty-two 12.7 cm cells. The spatial and energy resolution of the system remains sufficient to localize neutron and gamma sources, and to make spectral measurements that can distinguish among some source materials. As we show in section V, this configuration also leads to a compact overall system configuration by adding photomultiplier tubes to the tops or bottoms of the cells.

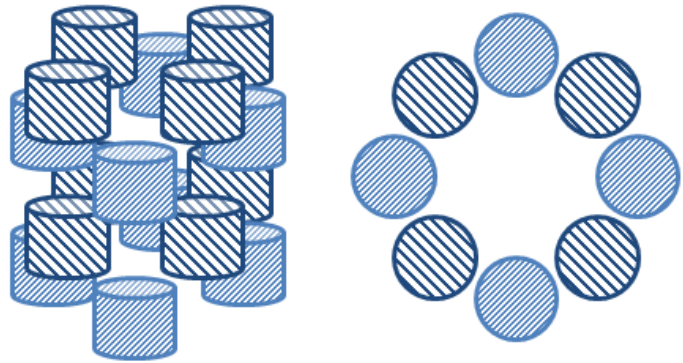

Fig. 3. Cell layout of the height offset cylindrical configuration of neutron detectors
(left: side view; right: top view)

## IV. Detector cell design and performance

We investigated the performance of a variety of 3"-diameter cells and photomultiplier combinations using Detect2000[18] to guide our experimental studies. After a variety of simulations and measurements of sensitivity and PSD performance (including investigating a variety of photomultiplier choices), we chose cylindrical, 7.6 cm diameter, 7.6 cm deep liquid scintillator cells coupled to a 7.6 cm diameter photomultiplier (Electron Tubes 9821B). Based on these results and previous experience with a wide variety of liquid scintillator cells, we have developed a new cell design explicitly for this project (Fig. 4). The body of each cell is constructed out of a solid billet of 6061-T6 aluminum. Aluminum offers a strong yet light material that is easy to machine and has an acceptable level of interaction with both neutrons and gamma rays. Cell interior walls are painted with EJ-520 to give a uniform and diffuse reflective surface. Each cell has a window at one end of the right cylinder that is made of cast acrylic as it is optically clear in the scintillators emission spectrum, has excellent machinability, is functional from -20 to 180 degrees F, and is highly impact and chemically resistant. The windows are coupled to the cell using a mechanical flange and sealed with Viton O-rings. Viton is used for superior environmental stability and chemical resistance. To enable a bubble free volume of EJ-309 scintillator, each cell has an expansion chamber to allow a useful temperature range of 50 degrees Celsius. The expansion chamber uses a small cylinder and piston coupled to the main volume via a small hole in the floor of the cell. The chamber piston uses a double O-ring seal in intimate contact with the liquid in the chamber. As the temperature changes, the expansion and contraction of the liquid creates a hydraulic force on the piston, moving it in and out. This movement compensates for the expansion and allows the cells to be virtually bubble free. Photomultiplier tubes are coupled to the cell window using silicone coupling material and held in place using light spring pressure. The entire PMT assembly (Fig. 4), including the divider base and high voltage power supply (a low-power, USB-controlled custom circuit mounted directly behind the voltage divider board), is housed in a light tight assembly lined with Mu-metal.

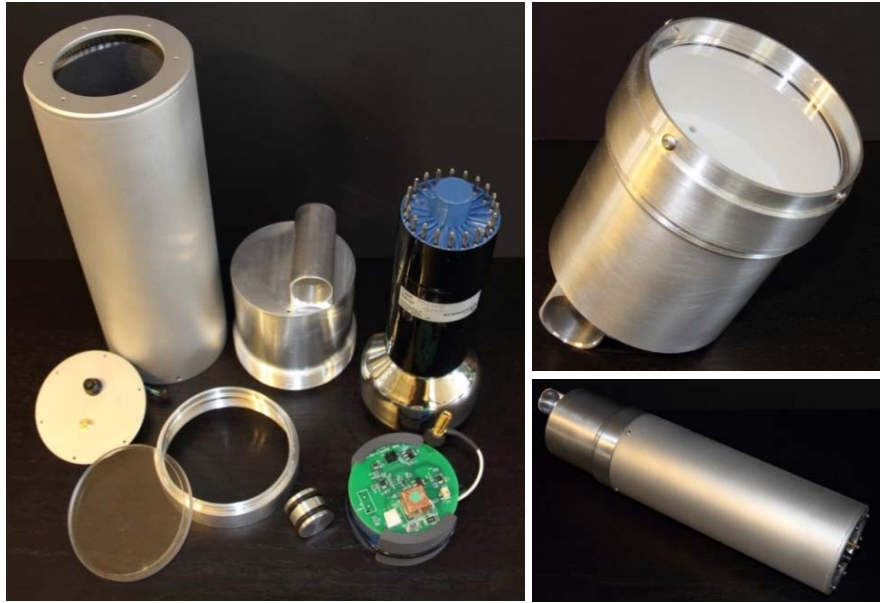

Fig. 4. Detector cell components (left), assembled liquid scintillator cell (top right), and complete detector assembly (bottom right).

The three key considerations related to the performance of the individual cells are our ability to measure the energy deposited in a cell when an incident neutron scatters off a proton in the scintillator ($E_{D1}$ in Figure 1), ability to discriminate neutrons from gammas based on the shape of the PMT output pulse, and a fast time response in order to accurately measure the time-of-flight τ. We measured these characteristics using a laptop computer to control a Struck SIS3316 16-channel, 250 MSPS, 14-bit digitizer. The digitizer firmware is very flexible, with multiple control and readout modalities. During normal operations, the digital signal processors in the module provide peak signal values, signal sums over specified windows, a timestamp that counts 4-ns clock periods since the start of data acquisition, and internal "moving average window" processing that can be used to calculate event timing with substantially better precision than the 4-ns-period sampling clock. These capabilities enable us to calculate all necessary quantities (scintillation signal strength, PSD, and time-of-flight values) on an event-by-event (real-time) basis. The digitizer can also provide full waveforms for more in-depth analysis.

Figure 5 shows a representative measurement of the peak amplitude spectrum of the gamma emissions from a $^{22}$Na source, which we use to calibrate the detector. We start with an MCNPX-PoliMi[19] model of the energy deposited by a $^{22}$Na source in our detector cell, adding Gaussian broadening and an exponential low-energy background to obtain a model detector response curve. We then fit this to the measurement, with the two Compton edges at 341 keV and 1062 keV making it possible to convert peak amplitude into an electron equivalent energy deposited. The agreement of the fit to the measurement indicates that the detectors are responding as expected.

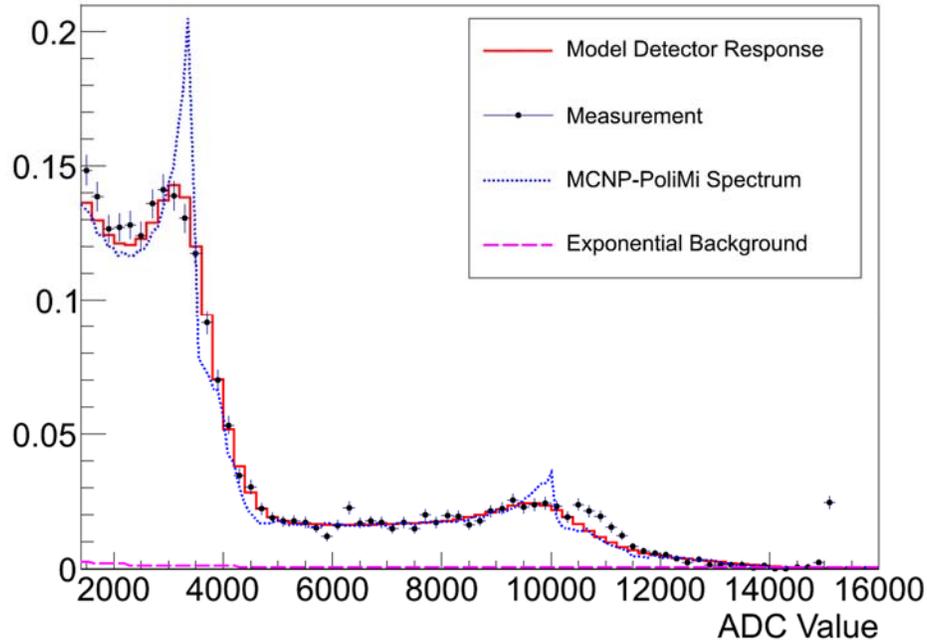

Fig. 5. Liquid scintillator cell energy calibration.

In order to distinguish neutron and gamma signals, we integrate the PMT output pulse over a long and a short time interval (using the windowing capability of the digitizer), and define a PSD response as the baseline-subtracted ratio of the long-interval pulse over the short-interval pulse. Because neutrons produce scintillation pulses with longer tails than gammas in the liquid organic scintillator, neutrons have larger PSD values than gammas. Each data point in Figure 6 represents the PSD value for individual scintillation events as a function of the peak amplitude of the corresponding scintillation pulse; neutrons appear in the upper branch of the figure (PSD>1.2), gammas in the lower branch (PSD<1.15). For a specified pulse peak amplitude (a vertical slice through the figure), we characterize the PSD response as a fit to a sum of two Gaussians. The fit then enables us to assign a statistical probability that each pulse was produced by a neutron or a gamma interaction (represented by colors and shading in the figure). The "staircase" appearance in the low-amplitude region of the plot is caused by the relatively coarse amplitude binning used in this analysis.

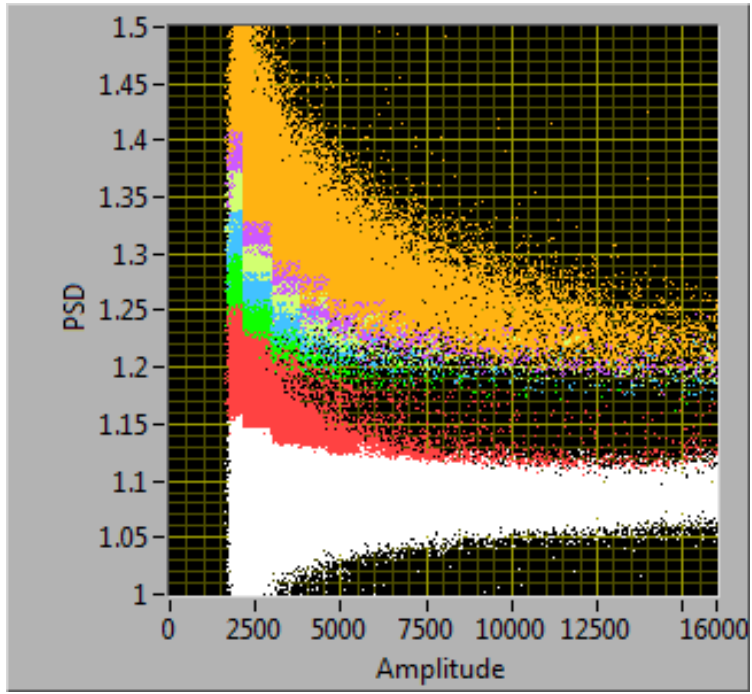

Fig. 6. Neutron-gamma discrimination using pulse-shape discrimination (PSD). White: gamma events (>90% probability). Red: identification uncertain. Other colors: neutron events (statistical certainty >90% for green, >99% for blue, >99.9% for yellow, >99.99% for magenta, >99.999% for orange).

For each digitizer event, in addition to the 4-ns resolution event timestamp, the SIS3316 digitizer also provides moving-average-window timing values that enable us to improve the timing resolution beyond that provided by the raw timestamp. In order to determine this accuracy, we placed a gamma source to one side of the system, and recorded the time difference between the trigger times for two detectors. Figure 7 displays a histogram of that time difference, with a FWHM of 1.25 ns for a Gaussian fitted to the result. To take into account differences in the time responses of the individual channels, we placed a gamma source at the center of the array of detectors, and then for each pair of detectors we recorded the difference in time between a gamma that is scattered from one to the other. We then used a least-squares fitting routine incorporating measurements from all combinations of two-cell timing signals to determine single-channel timing corrections to minimize this effect. This enables us to measure the time interval between double-scatter events with ~1-ns precision. This is sufficient for the neutron time-of-flight measurement, though it is marginal for determining the sequence of double-scatter events needed for gamma imaging.

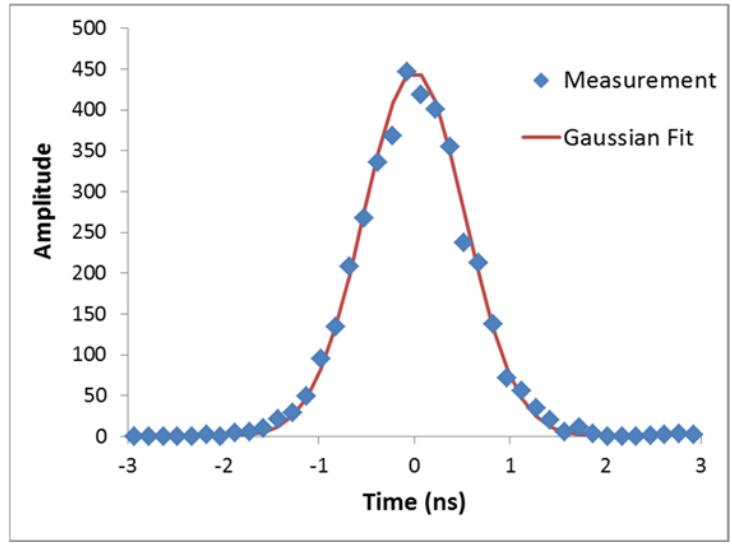

Fig. 7. Histogram of the variation in gamma transit time between two cells of the detection system

**V.   System implementation**

Fig. 8 shows the design of our final ruggedized system. The cells are suspended in the center of the external protective cylinder (0.9 m high, 0.4 m diameter), with shock mounting provided by the round structures at the top and bottom of the assemblies. The nearest-neighbor spacing between the detection cells is equal to their diameter and height (7.6 cm). The photograph in Fig. 9 shows the final assembly opened up to reveal its internal structure.

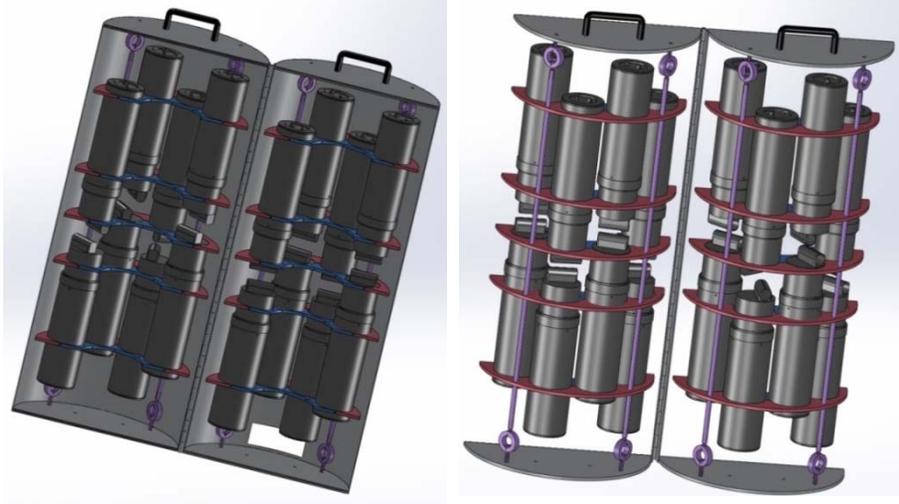

Fig. 8. Illustration of the mechanical design of the detection system.

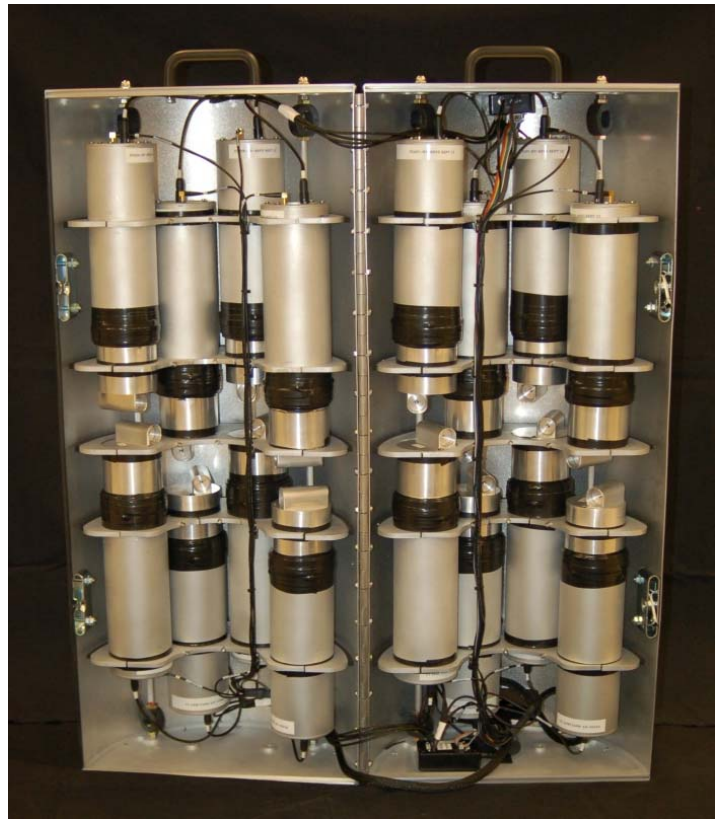

Fig. 9. Photograph of the interior of the detection system.

The initial implementation of the electronics used a relatively high-power VME (Versa Module Europa bus) based data acquisition system. Since that time, Struck has introduced a DC-powered "desktop" version of the sixteen-channel digitizer that can be operated using Ethernet, eliminating the need for any VME-based electronics. This new product has made it straightforward to modify the system so that it can be powered from either an electrical outlet or a battery.

The new desktop digitizer operates with 19-36 VDC power. The PMT circuitry operates on 12 VDC power (up-converted to high voltage using the custom circuitry mounted behind each PMT base shown in Fig. 4). To provide these two voltages, we have built a power conditioning module that can in turn be powered by electrical outlet or battery power (110 W required). We have operated the system for five hours using a nominally 92 amp-hour deep-cycle battery, observing a 1 V drop in battery voltage after that period, which corresponds to roughly half the nominal capacity of the battery. The digitizer draws ~90 W, and the PMTs draw ~20 W. The latter could be reduced substantially by replacing the standard resistor-based dynode chain with a transistorized circuit, but significant reductions in power can only be achieved by reducing the power required by the digitizer. The digitizer we are using was originally designed for use in a VME crate, and thus low power consumption was likely not an overriding concern. With careful attention to the power consumption of all components, we believe that the system power consumption could be reduced substantially.

To simplify transport and handling of the system, we chose to mount the electronics on the side of the cylindrical detector head (see Fig. 10). The top box is the power conditioning module described above, and the bottom module is the desktop digitizer. With this configuration, only three external cables are connected to the assembly: power (either AC power or battery), a USB cable to control the PMT high voltage, and an Ethernet cable to communicate with the digitizer. The latter two cables are connected to the Windows-based laptop computer shown resting on top of the system (the only other component of the entire system) that runs the LabVIEW-based data acquisition and real-time display software. For transport, it is only necessary to disconnect the three cables before the assembly and the laptop are placed in a 1.1 m x 0.7 m x 0.5 m Pelican transport case. The system can be set up at a new location and be ready to operate in less than five minutes. The energy calibration of the cells can be confirmed (and if necessary adjusted) by performing a several-minute measurement of a $^{60}$Co source placed on top of the cylinder. It is not necessarily to precisely match the calibration of the cells, as a calibration factor for each channel can be quickly determined from this measurement and entered into a system parameter file.

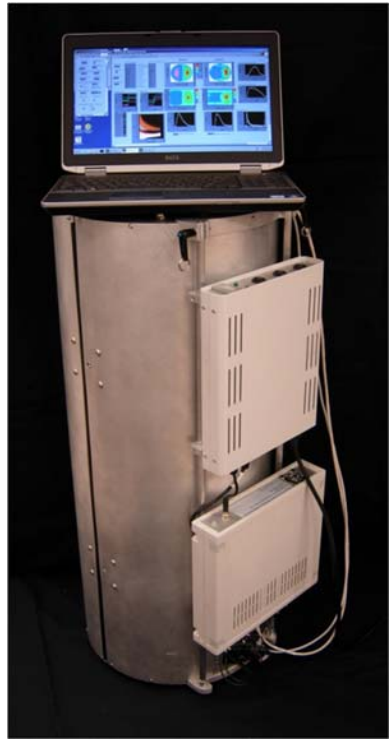

Fig. 10. Photograph of the detection system

One disadvantage of this mounting scheme is that it introduces additional material on one side of the detector head, leading to a small asymmetry in the response of the system in the horizontal plane. We evaluated this effect by measuring the response to a $^{252}$Cf source located 2 m from the system positioned directly in-line with the side-mounted modules, recording measurements both with the modules in place and with them removed. We found that the presence of the electronics reduced the neutron and gamma count rates by ~5% and ~10%, respectively. We did not observe any significant discrepancies between the resulting "images" or neutron spectra, however, and for most applications we believe that the asymmetry introduced by the presence of the electronics will not be significant. However, if desired, the electronics can be removed from the side of the detector head simply by pulling the two pins visible at the top of the module mounting rails, freeing the rails so that the modules can be relocated as a unit.

Alternatively, as it is unlikely that the source being sought would be directly above or below the system, the electronics could be mounted on its top or bottom (although the resulting height of the system would make it significantly more cumbersome).

The system is operated using LabVIEW-based software on a standard Windows laptop PC. Data acquisition cycles operate on a user-specified time period, during which one memory buffer of the digitizer acquires measurements while data is transferred from, stored to disk, and optionally processed and displayed from the other buffer. At the end of each period, the role of the two buffers switches, eliminating data-transfer deadtime (as long as the buffer can be read out faster than it is filled). The software can acquire and store full digitized waveforms, but it is more typically configured such that the digitizer only transfers key pulse parameters (pulse height, three integration gates used for PSD, and timing information), greatly reducing the data-transfer overhead. The real-time software display includes time-cycle-updated images, spectra, PSD characterization, and event counts (classified by type) at event rates in excess of $10^5$ events per second while also streaming the measurements to disk at the completion of each buffer read (list-mode recording). Operation at roughly two-times higher trigger rates can be accommodated by simply turning off the real-time display.

## VI.   System characterization

We have performed extensive laboratory characterization of the system using $^{252}$Cf and AmBe neutron sources and a variety of gamma sources. As one example, Fig. 11 presents a sample screenshot from the real-time display provided by the LabVIEW program. This twenty-minute measurement was recorded with the sources placed only ~2 m from the detector in order to provide very clean images and spectra. The screen is updated during the course of a measurement at a user-specified rate, typically 10 seconds for strong nearby sources to a minute or two for weaker more distant sources. The "PSD Graph" below the table is a real-time display of those events (see Fig. 6). The "Neutron TOF" graph below it is a histogram of the neutron transit time between the two scattering cells ($\tau$ in Eqn. 1). In the "Neutron" column in Fig. 10, the top plot is a real-time image of the neutron source created as a back-projection reconstruction of events using Eqn. 3 (we have also implemented post-data-acquisition image generation using maximum likelihood expectation maximization, or MLEM, and hypothesis-test methods). The x-axis, $\theta$, is equivalent to latitude (-180º – +180º), and the y-axis, $\varphi$, is equivalent to longitude (+90º straight up, -90º straight down). The bottom graph is an energy spectrum (in MeV) constructed using the event-by-event energy values derived using Eqn. 2; longer data acquisition times are necessary to obtain useful spectra. The "number of neutron pairs" value at the bottom of the column is the number of double-scatter events that were used to derive the information described above. The data in the "Gamma" column is similar. Additional diagnostic plots can also be displayed (histograms of the neutron and gamma double-scatter angles and the neutron and gamma scintillator signals recorded in all cells). This same routine is used for post-acquisition analysis of measurements saved to disk, with the ability to vary analysis parameters at that time.

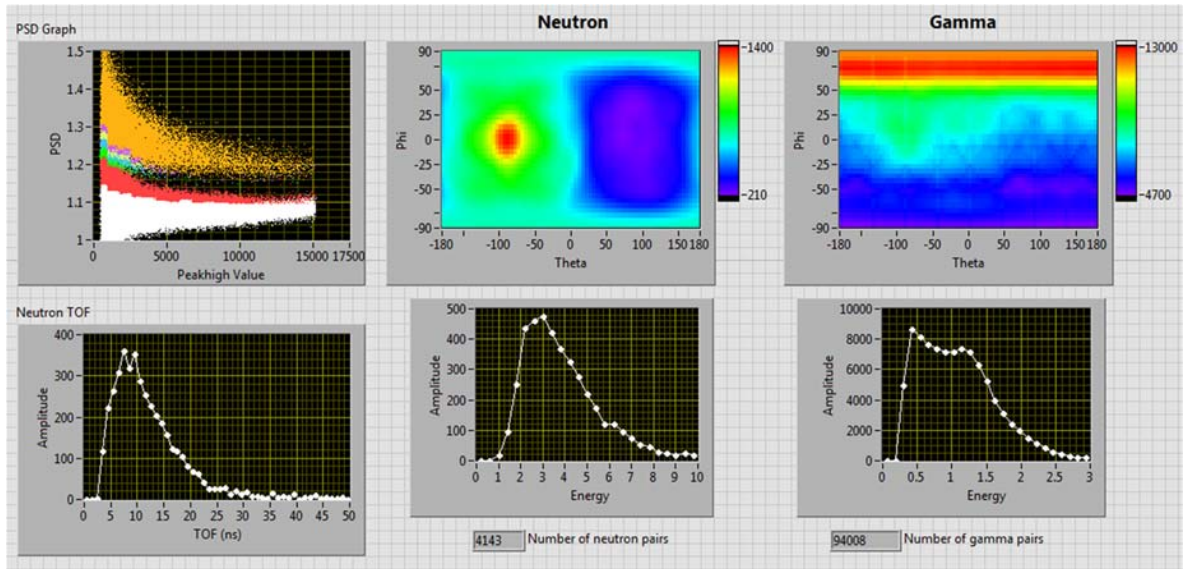

Fig. 11. Real-time display at the end of a measurement made with a $^{252}$Cf placed at θ = -90º, φ = 0º and a $^{60}$Co gamma source placed at φ = 90º directly above the detector.

In order to be designated a double-scatter neutron event, we require that the pulse in each single-cell be categorized as a neutron event by PSD, and that the time difference between the two events is longer than it would be for a gamma. This triple requirement makes the neutron imaging very robust in the presence of strong gamma fields. This is demonstrated in Fig. 12, in which the neutron image of a $^{252}$Cf source is unaffected by the presence of the strong gamma emission from a $^{152}$Eu source. (The $^{252}$Cf gamma image is not visible in the $^{252}$Cf+$^{152}$Eu gamma image because its rate of valid double-scatter events is nearly 100 times lower than that of $^{152}$Eu.)

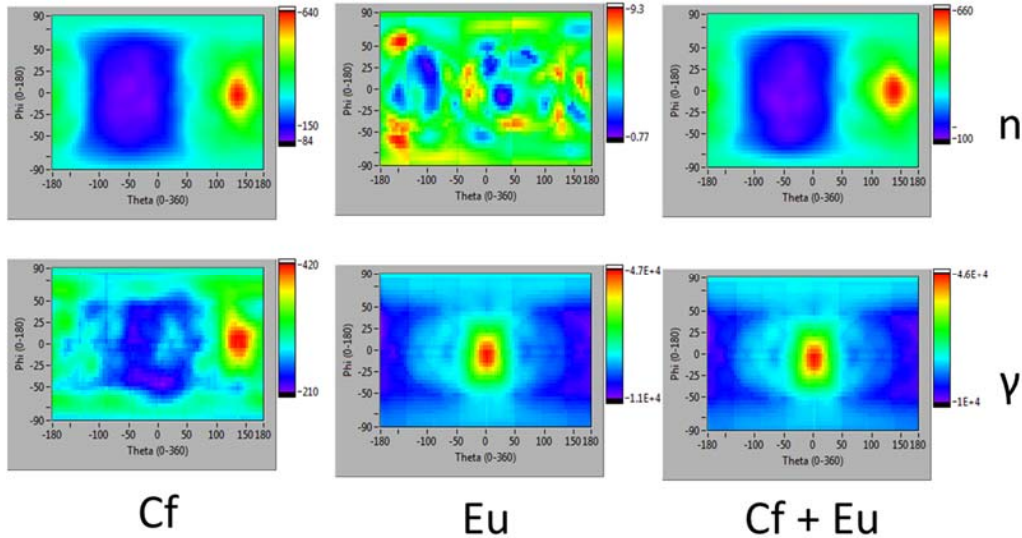

Fig. 12. Neutron (top) and gamma (bottom) images recorded with a $^{252}$Cf source several meters from the system at a relative angle of 135º (left), a strong $^{152}$Eu source closer to the system at a relative angle of 0º (center), and both sources at the same time (right). The scales on the color bars indicate the relative strengths of the signals (in particular, no neutron signal for $^{152}$Eu only).

To characterize the intrinsic efficiency of the system for detecting neutrons, we consider the cross sectional area of the 32 cm wide by 30 cm high rectangle of the sixteen liquid scintillator cells viewed from the side of the apparatus (ignoring for this purpose the majority of the system area/volume taken up by the photomultipliers and their associated hardware). With this definition, and using $^{252}$Cf with known neutron emission rates, we measure a 45% efficiency for detecting fission-energy neutrons that strike that area, and a 1% efficiency for observing double-scatter events for fission-energy neutrons that strike that area (single-cell neutron energy threshold ~0.5 MeV). To partially characterize the angular resolution of the system, we note that we measure a typical standard deviation of ~25° in the scattering angle for single pairs of detection cells (i.e. multiple measurements of the angle θ shown in Fig. 1). A full discussion of the resulting image resolution is beyond the scope of this paper.

## VII. Neutron spectroscopy

Detailed measurements of three test objects that contain significant quantities of plutonium were made during a measurement at the Idaho National Laboratory ZPPR (Zero Power Physics Reactor) facility. One object is a plutonium metal sphere; the other two are "hex cans" containing plutonium oxide. In addition to performing a variety of imaging studies, the relatively high neutron emission rates of these objects provided excellent statistics for measurements of neutron energy spectra. The plutonium metal sphere emits only fission-spectrum neutrons, while the hex cans additionally emit neutrons from an (α,n) interaction on oxygen in the material. The differences in these two neutron spectra can be used to distinguish plutonium metal from its oxide.[20-24] Measurements of double-scatter and single-scatter events are shown in Figs. 13 and 14, respectively, demonstrating the effectiveness of the detector system for making this material discrimination. Spectra are also displayed for $^{252}$Cf (a fission-energy spectrum that for all practical purposes cannot be distinguished from plutonium metal) and AmBe (a common industrial neutron source). These spectra do not represent the true spectra, as the instrument response function has a steep threshold at low-energies. However, the differences between observed spectra can nonetheless be used for material characterization.

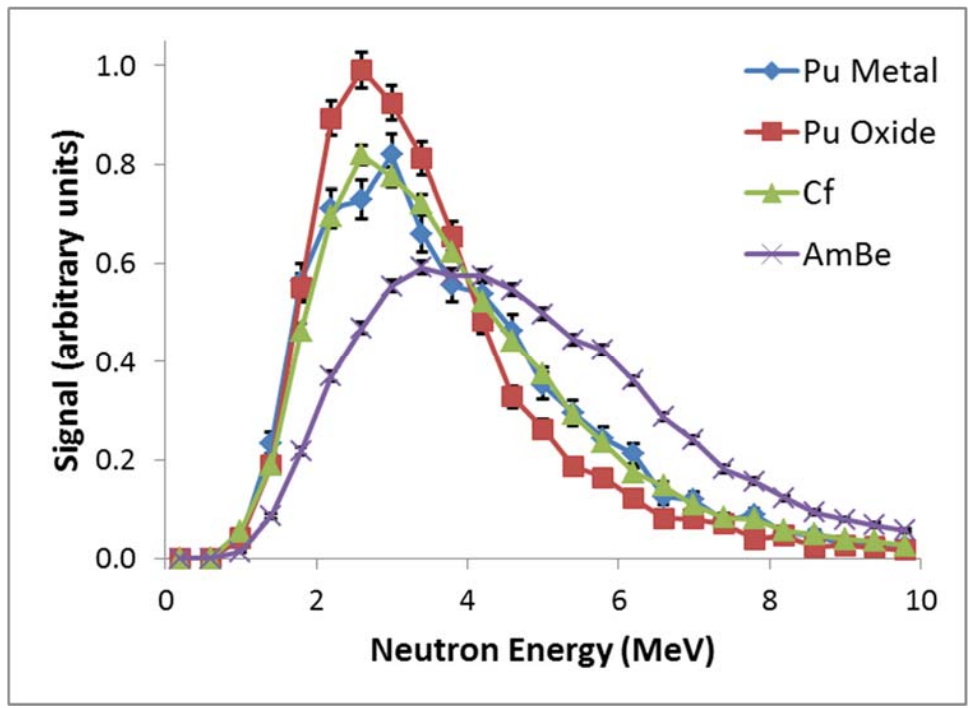

Fig. 13. Area-normalized double-scatter spectra (not corrected for instrument response) for four neutron sources. The error bars represent one standard deviation statistical error, assuming a Poisson distribution in the number of events in each histogram bin.

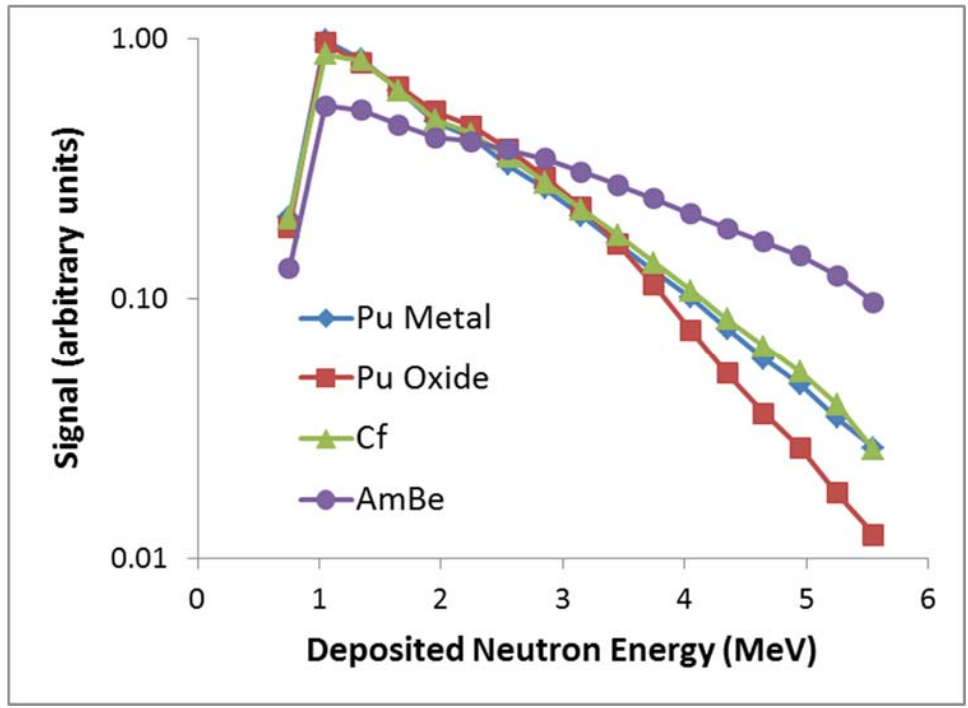

Fig. 14. Area-normalized single-scatter spectra (not corrected for instrument response) for four neutron sources. The statistical errors are less than or on the order of the size of the plot symbols.

The spectra shown in Figs. 13 and 14 are for bare sources, whereas material of interest in real-world scenarios is likely to be shielded by surrounding material. To explore this impact on fission-energy neutron spectra, peak-normalized double-scatter and single-scatter spectra are displayed for a bare $^{252}$Cf source, and the same source shielded by 6" of high-density polyethylene (HDPE) and 2" of lead (Figs. 15 and 16). Although the intervening material reduces the strength of the signal, it has little impact on the shape of the recorded spectrum. This behavior (for double-scatter spectra in particular) can be attributed to the relatively high energy threshold of the detection system, the significant energy loss of the neutrons during scattering events, and the fact that a significant fraction of fission-energy neutrons fall below the energy threshold.

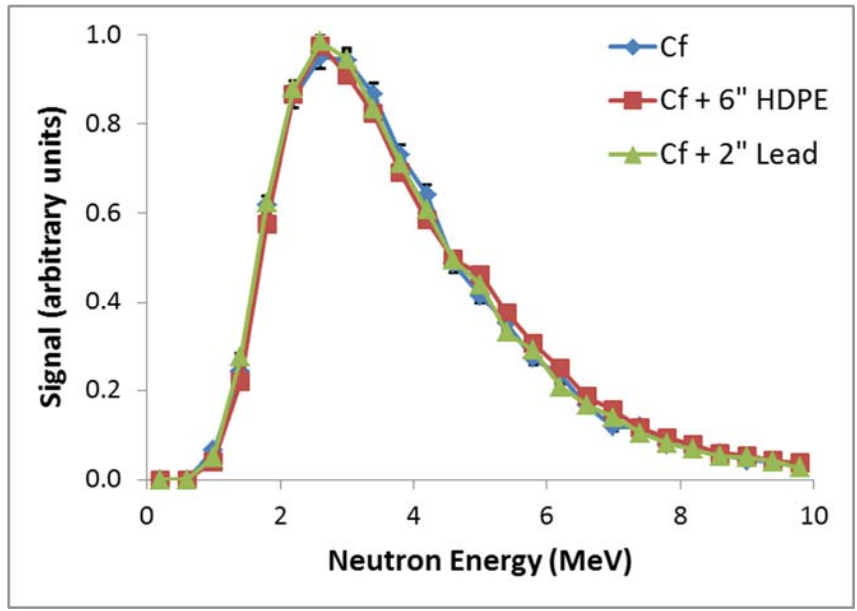

Fig. 15. Area-normalized double-scatter spectra (not corrected for instrument response) for a bare and shielded $^{252}$Cf source. The error bars (barely visible) represent one standard deviation statistical error, assuming a Poisson distribution in the number of events in each histogram bin.

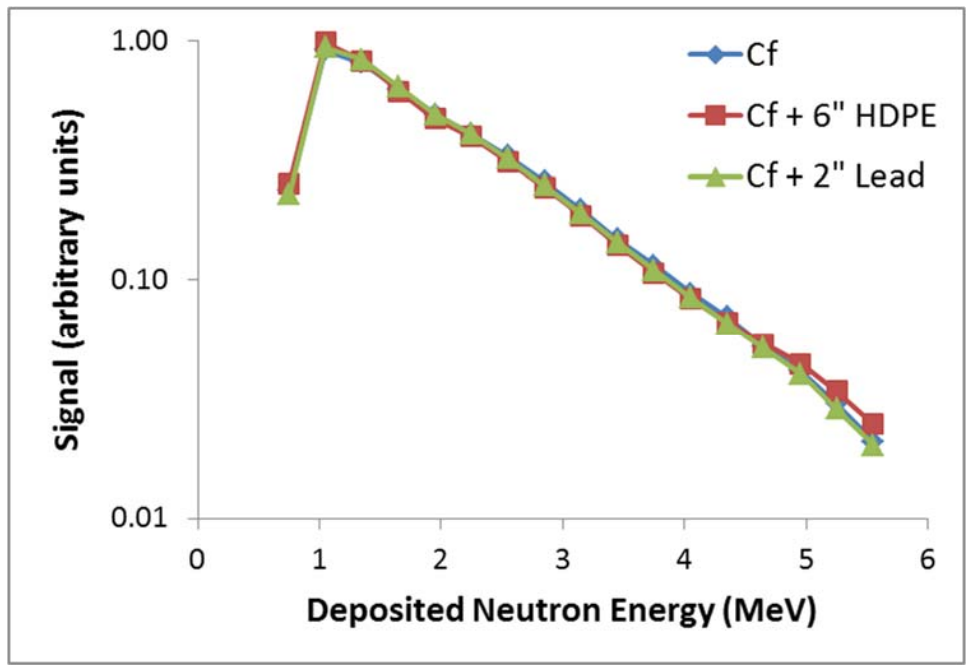

Fig. 16. Area-normalized single-scatter spectra (not corrected for instrument response) for a bare and shielded $^{252}$Cf source. The statistical errors are less than or on the order of the size of the plot symbols.

## VIII. Conclusion

We have described the design and performance of a compact, omnidirectional-sensitivity neutron scatter camera, touched on its capabilities for localizing both neutron and gamma sources, and demonstrated its use for material identification using neutron spectroscopy. Although not described here, this system has been very successful at localizing and characterizing neutron (and to a lesser extent gamma) sources during several field deployments. Future work will explore detection of HEU through active interrogation, neutron source multiplication measurements using time-correlated pulse-height measurements,[25] enhanced gamma localization using energy windowing, and additional field deployments. We are also exploring the use of alternative detection materials to enhance the sensitivity and neutron-gamma discrimination capabilities of the system, and alternative photodetectors to substantially decrease its size.

**Acknowledgments**

We thank Nick Mascarenhas for initiating this project, Scott Kiff for his support as interim principal investigator, Peter Marleau and Erik Brubaker for sharing their expertise about neutron scatter cameras, and Patricia Schuster (University of California at Berkeley) for her work modeling a variety of cell configurations. This work is supported by the National Nuclear Security Administration Office of Defense Nuclear Nonproliferation Research and Development, Nuclear Weapon and Material Security Team. Sandia National Laboratories is a multi-program laboratory managed and operated by Sandia Corporation, a wholly owned subsidiary of Lockheed Martin Corporation, for the U.S. Department of Energy's National Nuclear Security Administration under contract DE-AC04-94AL85000.

**References**

1. N. Mascarenhas, J. Brennan, K. Krenz, J. Lund, P. Marleau, J. Rasmussen, J. Ryan, and J. Macri, "Development of a Neutron Scatter Camera for Fission Neutrons," 2006 IEEE Nuclear Science Symposium Conference Record, San Diego, CA, 185 (2006).

2. N. Mascarenhas, J. Brennan, K. Krenz, P. Marleau, and S. Mrowka, "Results with the neutron scatter camera," IEEE Transactions on Nuclear Science **56**, 1269 (2009).

3. J. Brennan, E. Brubaker, R. Cooper, M. Gerling, C. Greenberg, P. Marleau, N. Mascarenhas, and S. Mrowka, S., "Measurement of the fast neutron energy spectrum of an $^{241}$Am-Be source using a neutron scatter camera," IEEE Transactions on Nuclear Science **58**, 2426 (2011).

4. M. D. Gerling, J. E. M. Goldsmith, and J. S. Brennan, "MINER – a mobile imager of neutrons for emergency responders," 2014 IEEE Nuclear Science Symposium and Medical Imaging Conference Record (NSS/MIC), Seattle, WA (2014).

5. R. S. Miller, J. R. Macri, M. L. McConnell, J. M. Ryan, E. Flückiger, and L. Desorgher, "SONTRAC: An imaging spectrometer for MeV neutrons," Nuclear Instruments and Methods in Physics Research A **505**, 36 (2003).

6. P. E. Vanier and L. Forman, "An 8-element neutron double-scatter directional detector," Proc. SPIE 5923, Penetrating Radiation Systems and Applications VII, 592307 (2005).

7. P. E. Vanier, L. Forman, C. Salwen and I. Dioszegi, "Design of a large-area fast neutron directional detector," 2006 IEEE Nuclear Science Symposium Conference Record, San Diego, CA, 93 (2006).

8. U. Bravar, P. J. Bruillard, E. O. Fluckiger, J. R. Macri, and A. L. MacKinnon, "Development of the fast neutron imaging telescope," IEEE Nuclear Science Symposium Conference Record **1**, 107 (2005).

9. U. Bravar, P. J. Bruillard, E. O. Fluckiger, J. Legere, and J. R. Macri, "Design optimization and performance capabilities of the fast neutron imaging telescope (FNIT), IEEE Nuclear Science Symposium Conference Record **1**, 264 (2007).

10. A. C. Madden, P. Bloser, D. Fourguette, L. Larocque, J. S. Legere, M. Lewis, M. L. McConnell, M. Rousseau, and J. M. Ryan, "An imaging neutron/gamma-ray spectrometer," Proc. SPIE **8710**, 87101L (2013).

11. A. Poitrasson-Rivière, M. C. Hamel, J. K. Polack, M. Flaska, S. D. Clarke, and S. A. Pozzi, "Dual-particle imaging system based on simultaneous detection of photon and neutron collision events," Nuclear Instruments and Methods in Physics Research A **760**, 40 (2014).

12. J. Beaumont, M. P. Mellor, and M. J. Joyce, "The analysis of complex mixed-radiation fields using near real-time imaging," Radiation Protection Dosimetry **161**, 331 (2014).

13. P. Hausladen, M. Blackston, E. Brubaker, D. Chichester, P. Marleau, and J. Newby, "Fast-neutron coded-aperture imaging of special nuclear material configurations," INMM Conference Proceedings (2012).

14. J. Brennan, E. Brubaker, P. Marleau, A. Nowack, and P. Schuster, "Results from field tests of the one-dimensional Time-Encoded Imaging System," Sandia Report SAND2014-17691 (2014).

15. J. Brennan, E. Brubaker, M. Gerling, P. Marleau, K. McMillan, A. Nowack, N. Renard-Le Galloudec, and M. Sweany, "Demonstration of two-dimensional time-encoded imaging of fast neutrons," Nuclear Instruments and Methods in Physics Research A **802**, 76 (2015).

16. Glenn F. Knoll, *Radiation Detection and Measurement*, 4th ed., Wiley, NY, 2010, p. 570, eqn. 15.3, with A=1 for proton target, and θ redefined as angle of scattered neutron from the direction of the incoming neutron (complement of θ used in eqn. 15.3).

17. Glenn F. Knoll, ibid, p. 324, eqn. 10.2 (solved for θ).

18. F. Cayouette, D. Laurendeau, and C. Moisan, Proc. SPIE 4833, "DETECT2000: an improved Monte-Carlo simulator for the computer aided design of photon sensing devices," Applications of Photonic Technology **5**, 69 (2003).

19. S. A. Pozzi, S. D. Clarke, W. J. Walsh, E. C. Miller, J. L. Dolan, M. Flaska, B. M. Wieger, A. Enqvist, E. Padovani, J. K. Mattingly, D. L. Chichester, and P. Peerani, "MCNPX-PoliMi for nuclear nonproliferation applications," Nuclear Instruments and Methods in Physics Research A **694**, 119 (2012).

20. J. l. Dolan, M. Flaska, S. A. Pozzi, and D. Chichester, "Measurement and Characterization of Nuclear Material at Idaho National Laboratory," Institute of Nuclear Materials Management 50th Annual Meeting, Tucson, AZ, July 12-16, 2009, and Report INL/CON-09-16103, Idaho National Laboratory, Idaho Falls, Idaho (2009).

21. S. A. Pozzi, S. D. Clarke, M. Flaska, and P. Peerani, "Pulse height distributions of neutrons and gamma rays for plutonium-oxide samples," Nuclear Instruments and Methods in Physics Research A **608**, 310 (2009).
22. J. M. Verbeke and G. F. Chapline, "Distinguishing Plutonium Metal from Plutonium Oxide using Fast Neutrons, Preliminary Results," LLNL-TR-599212 (2012).
23. J. M. Verbake, G. C. Chapline, L. F. Nakae, and S. A. Sheets, "Distinguishing Pu Metal from Pu Oxide using Fast Neutron Counting," Palm Desert, CA, United States (July 14-18, 2013) and LLNL-CONF-637437.
24. S. A. Pozzi, M. M. Bourne, J. L. Dolan, K. Polack, C. Lawrence, M. Flaska, S. D. Clarke, A. Tomanin, and P. Peerani, "Plutonium metal vs. oxide determination with the pulse-shape-discrimination-capable plastic scintillator EJ-299-33," Nuclear Instruments and Methods in Physics Research A **767**, 188 (2014).
25. Eric C. Miller, "Characterization of Fissionable Material using a Time-Correlated Pulse-Height Technique for Liquid Scintillators," Ph.D. thesis, University of Michigan, 2012, http://dnng.engin.umich.edu/wp-content/uploads/sites/101/2014/10/ericcm_1.pdf ).